\RequirePackage{fix-cm}
\documentclass[twocolumn]{svjour3}          

\smartqed  
\usepackage{mathptmx}      
\journalname{SOLARNET}
\usepackage{amsmath,amssymb}
\usepackage{hyperref}
\usepackage[]{graphicx}
\usepackage{multirow}
\usepackage{pdflscape}
\usepackage{array}

\begin{document}

\title{JP3D compression of solar data-cubes: photospheric imaging and spectropolarimetry}

\titlerunning{JP3D compression of solar data-cubes: photospheric imaging and spectropolarimetry}  

\author{Dario Del Moro \and
        Luca Giovannelli \and
        Ermanno Pietropaolo \and
        Francesco Berrilli 
}

\institute{D. Del Moro \at
              University of Rome ``Tor Vergata'', Department of Physics, Via Della Ricerca Scientifica 1, 00133 Roma, Italy\\
              \email{delmoro@roma2.infn.it}           
           \and
           L. Giovannelli \at
              University of Rome ``Tor Vergata'', Department of Physics, Via Della Ricerca Scientifica 1, 00133 Roma, Italy\\
           \and
            E. Pietropaolo \at  
              University of L'Aquila, Department of Physics and Chemistry Sciences, Via Vetoio, 67100 Coppito (AQ), Italy\\
           \and
             F. Berrilli \at
              University of Rome ``Tor Vergata'', Department of Physics, Via Della Ricerca Scientifica 1, 00133 Roma, Italy\\ 
             }

\date{Received: date / Accepted: date}


\maketitle

\begin{abstract}
Hyperspectral imaging is an ubiquitous technique in solar physics observations and the recent advances in solar instrumentation enabled us to acquire and record data at an unprecedented rate.
The huge amount of data which will be archived in the upcoming solar observatories press us to compress the data in order to reduce the storage space and transfer times.\\
The correlation present over all dimensions, spatial, temporal and spectral, of solar data-sets suggests the use of a 3D base wavelet decomposition, to achieve higher compression rates.
In this work, we evaluate the performance of the recent JPEG2000 Part 10 standard, known as JP3D, for the lossless compression of several types of solar data-cubes.
We explore the differences in:
a) The compressibility of broad-band or narrow-band time-sequence; I or V stokes profiles in spectropolarimetric data-sets; 
b) Compressing data in [x,y,$\lambda$] packages at different times or data in [x,y,t] packages of different wavelength; 
c) Compressing a single large data-cube or several smaller data-cubes; 
d) Compressing data which is under-sampled or super-sampled with respect to the diffraction cut-off.
\end{abstract}

%
%

\section{Introduction}
\label{sect:intro}  
The advances in instrumentation and acquisition systems enable the solar physics community to acquire data to an unprecedented rate. 
Both the advent of the present satellite solar observatories (e.g.: SDO \cite{SDO}, HINODE \cite{hinode}) and of the modern imaging instruments (e.g.: ROSA \cite{ROSA}, CRISP \cite{CRISP}, IBIS \cite{IBIS}) have pushed the capabilities of storage and transmitting data to new limits.
In the next years, the new instrumentation at the foci of the 4m class solar telescopes (DKIST \cite{DKIST}, EST \cite{EST}) will stream data at a $\sim1$GB/s rate. 
This rate and the typical 9h duration of the observation runs will very probably exceed the storage and the transmission capacity of even the largest science facilities.
This amount of data should be compressed and made available to download within the next observation run (i.e.: within $\sim24$h) at the observatories (presumably Hawaii and Canary islands).
Then the data will be downloaded and decompressed by the users around the world.
The solar physics community is therefore about to enter the challenge, already tackled by the computer science community, of handling extremely large amount of data and the resulting problems of data compression, with or without information loss.
In this paper, we introduce two typical solar data-sets to be used as paragon data-sets to perform tests on the performance of different compression algorithm and strategies.
We apply an established compression method, namely JP3D: the Part 10 of the JPEG2000 standard \cite{JPEG2000,schelkens06,bruylants07a}, to the data-sets and evaluate the computational costs and the benefits of such a compression algorithm for the solar data-set case.
Thus far, the JP3D standard has been mainly used for the compression of 3D medical imagery \cite{bruylants07b,kimpe07}.
Those and other studies have already evaluated the performances of JP3D on hyperspectral data-sets (e.g.: AVIRIS radiance data-set \cite{AVIRIS1,AVIRIS2}, SAR data \cite{SAR}, etc.), pointing out that its performance is strongly dependent on the characteristics of the compressed data-set.
Although the compression is made only once, the velocity and efficiency of the compression are at present the driver of our analysis.
The objective of this paper is therefore to evaluate JP3D for the compression of hyperspectral solar imagery in those cases where lossless compression is mandatory.
\section{JP3D from OpenJPEG}
\label{sect:JP3D}
Hyperspectral imaging is an ubiquitous technique in solar physics observations. 
Present day compression of solar data-set is customarily performed on 2D layers of the usually 4D data-sets (x,y,$\lambda$,t).
Given that hyperspectral imagery can also be considered a volumetric form, it is natural to consider JP3D for hyperspectral compression \cite{zhang09}.
More, the correlation present in hyperspectral data-sets over all dimensions, spatial, temporal and spectral, suggests the use of a 3D base wavelet decomposition, to achieve higher compressions.\\
We choose the JP3D compression for this analysis because it is efficient, open-source and well documented. 
OpenJPEG is an open-source JPEG2000 codec written in C language.
It has been developed in order to promote the use of JPEG 2000, a still-image compression standard from the Joint Photographic Experts Group (JPEG).
Since May 2015, it is officially recognized by ISO/IEC and ITU-T as a JPEG2000 Reference Software. 
Anyone can use the code. The only restriction is to retain the copyright in the sources or in the binaries documentation.
Also, the use of the OpenJPEG implementation allowed us to compare the performance of JP3D versus the performance of the JPEG2000 performed in a previous study \cite{delmoro11}.

\section{The data-sets}
\label{sect:data}
The data-sets used in this work are a partial description of the present variety of high resolution photospheric solar data, which will probably be maintained in the next years. 
Data-sets include calibrated broad band, narrow band and spectropolarimetric observations of the photosphere.
They also represent different typologies of targets: the 2006 data-set images a Quiet Sun region at disk centre; the 2008 data-set images a small pore near disk centre.
Therefore, they allow us to evaluate the performance of the compressing algorithm in two different, but 'typical' situation of solar observation: low-contrast images with localized small-scale magnetic fields and high contrast and large and diffuse magnetic field.\\
The two data-sets are retrievable from \url{https://www.fisica.uniroma2.it/~solare/en/?p=257} to allow anyone to compare the compression performances of different algorithms on the same data-sets.\\

\subsection{2006 data-set} 
\label{sect:2006}
   \begin{figure*}
   \begin{center}
   \begin{tabular}{c}
   \includegraphics[width=12cm]{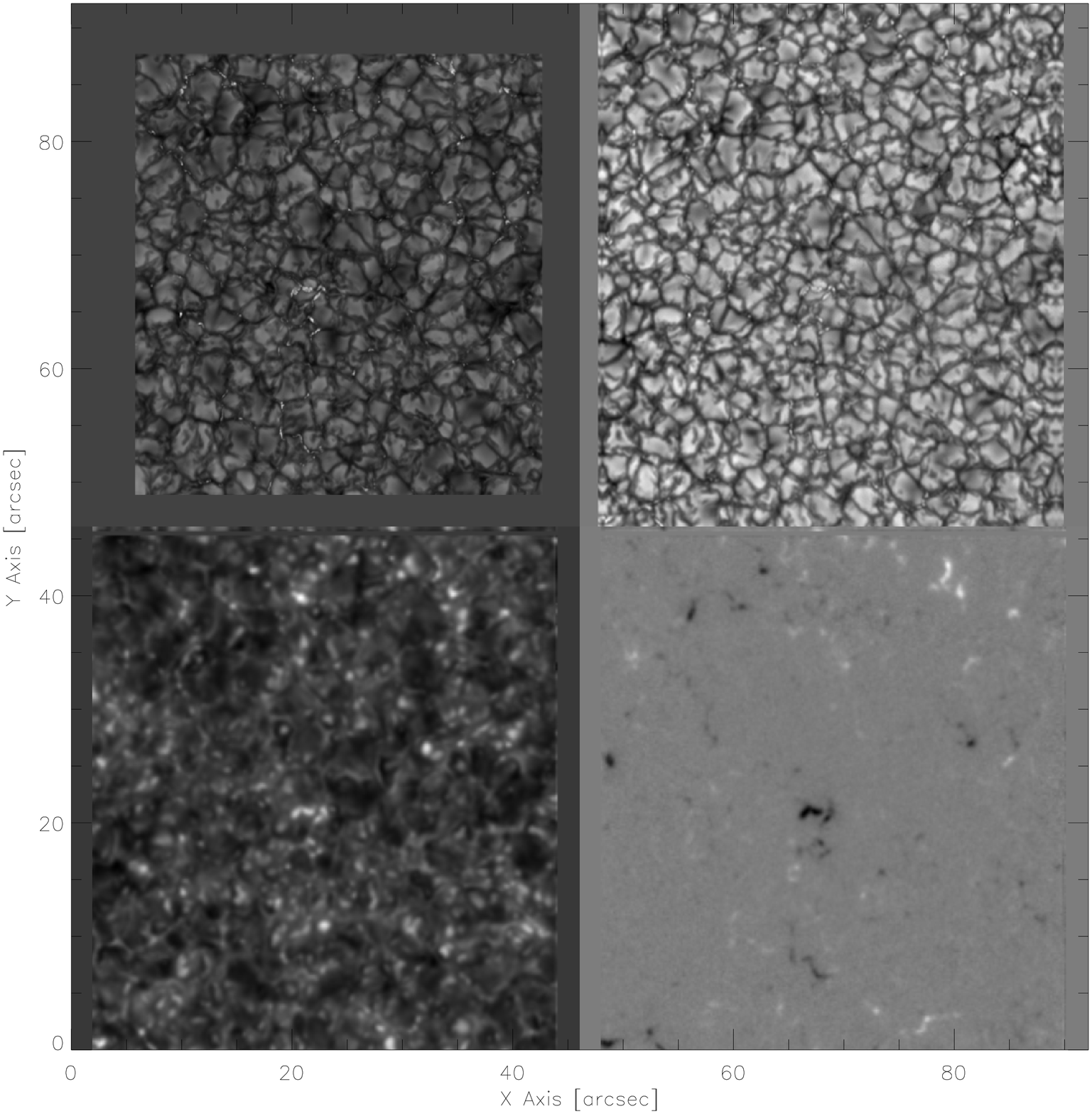}
   \end{tabular}
   \end{center}
   \caption 
   { \label{fig:2006} 
Some sample images from the 2006 data-set data-cubes. Upper-left: G-band image; Upper-right: Broad-band image; Lower-left: Stokes-I image near the core of the FeI 630.2 nm line; Lower-right: Stokes-V image near the wing of the FeI 630.2 nm line. For the sake of visualization, all images have been rescaled to the same pixel resolution (0.18" pixel$^{-1}$) and their values have been separately linearly scaled to saturate the greyscale palette. } 
   \end{figure*} 
The first data-set used \cite{bart2010} was observed on November 21, 2006 at the NSO/Dunn Solar Telescope (DST \cite{DST}), which has an effective entrance pupil of $\simeq0.76$~m, imaging a quiet region at disk center.
The field-of-view (FOV) was approximately $40''\times40''$.\\
The spectropolarimetric part of the data-set consists of $50$ Full-Stokes scans of the two FeI~$630$~nm lines performed by the Interferometric BIdimensional Spectrometer (IBIS), with a cadence of $89$ seconds.
The lines were sampled at $45$ wavelength points with a spectral FWHM of $2$~pm	and a step of $2.3$~pm.
The spectropolarimetric measurements, at each wavelength position are obtained as  I+S and I-S images on the same chip, with the following temporal scheme: S = [+V , -V , +Q, -Q, +U, -U].
These spectropolarimetric images have a pixel scale of $0.18''$.\\
Simultaneous to narrow-band spectropolarimetric data, broad-band ($633.32\pm5$~nm) counterparts of the same FOV were acquired (a sample of broad-band image shown in the upper-right panel of Fig. \ref{fig:2006}).
The exposure time for both narrow-band and broad-band data is $80$~ms.
The original pixel scale of the broad-band data is $0.09''$.\\
As a first step of the calibration procedure, the Multi-Frame Blind Deconvolution algorithm (MFBD\cite{MFBD}) has been applied to the broad-band images to obtain a master frame from each scan which had much reduced seeing degradation and a homogeneous resolution in the whole FOV.
We then applied a de-stretch process on the single broad-band images, using this master frame as reference image, then re-binning the broad-band images at the pixel scale of the spectropolarimetric images.
The computed de-stretch matrices were successively applied also onto the spectropolarimetric images.
With this operation, the quality of spectropolarimetric data has been improved, since the seeing-induced cross-talk was greatly reduced. 
The angular resolution of the spectropolarimetric data-set has been estimated to be $0.4''\simeq2$~pixels, which is close to that of the individual narrow band images.
The original IBIS spectropolarimetric $I \pm S$ images were successively combined and corrected for instrumental polarization to retrieve the Stokes vectors (samples of Stokes I and Stokes V images are shown in the lower-left and lower-right panels of Fig. \ref{fig:2006}, respectively).
The average noise level for Stokes $V$, measured as standard deviation of the circular polarization in continuum wavelengths, is $\sigma_V=3\cdot10^{-3} I_C$ (here $I_C$ is the continuum intensity).\\
To complement the broad-band and narrow-band data, G-band filtergrams ($430.5\pm0.5$~nm) were acquired for a slightly smaller FOV; the original pixel scale of such filtergrams was $0.037''$, while the exposure time was $15$~ms.
A post processing MFBD procedure was applied also to the G-band images, which were afterward re-binned at $4\times$ the spectropolarimetric spatial scale (a sample is shown in the upper-left panel of Fig. \ref{fig:2006}).\\
For the compression analysis, the spectropolarimetric and broad-band image were cropped to their $256 \times 256$ central parts and the G-band filtergrams were instead framed into $1024 \times 1024$ images (as shown on Fig. \ref{fig:2006} and resumed in Tab. \ref{tab:2006}).
All the data-cubes were saved as 32bit greyscale.\\
\begin{table*}[h]
\begin{small}
\caption{2006 data-set description}
\label{tab:2006}       
\begin{tabular}{lccccc}
\hline\noalign{\smallskip}
Type & Image Size & N. of frames & Repetitions & Spatial Resolution & Pixel Scale\\ 
\noalign{\smallskip}\hline\noalign{\smallskip}
G-band & 1024$\times$1024 & 18  & 50 & $0.16''$ & $0.05''$\\ 
\hline 
Broad-band & 256$\times$256 & 270 &  50 & $0.40''$ & $0.18''$\\ 
\hline 
Stokes I & 256$\times$256 & 45 & 50 & $0.40''$ & $0.18''$\\ 
\hline 
Stokes V & 256$\times$256 & 45 & 50 & $0.40''$ & $0.18''$\\ 
\noalign{\smallskip}\hline
\end{tabular}
\end{small}
\end{table*}

\subsection{2008 data-set}
\label{sect:2008}
   \begin{figure*}
   \begin{center}
   \begin{tabular}{c}
  \includegraphics[width=12cm]{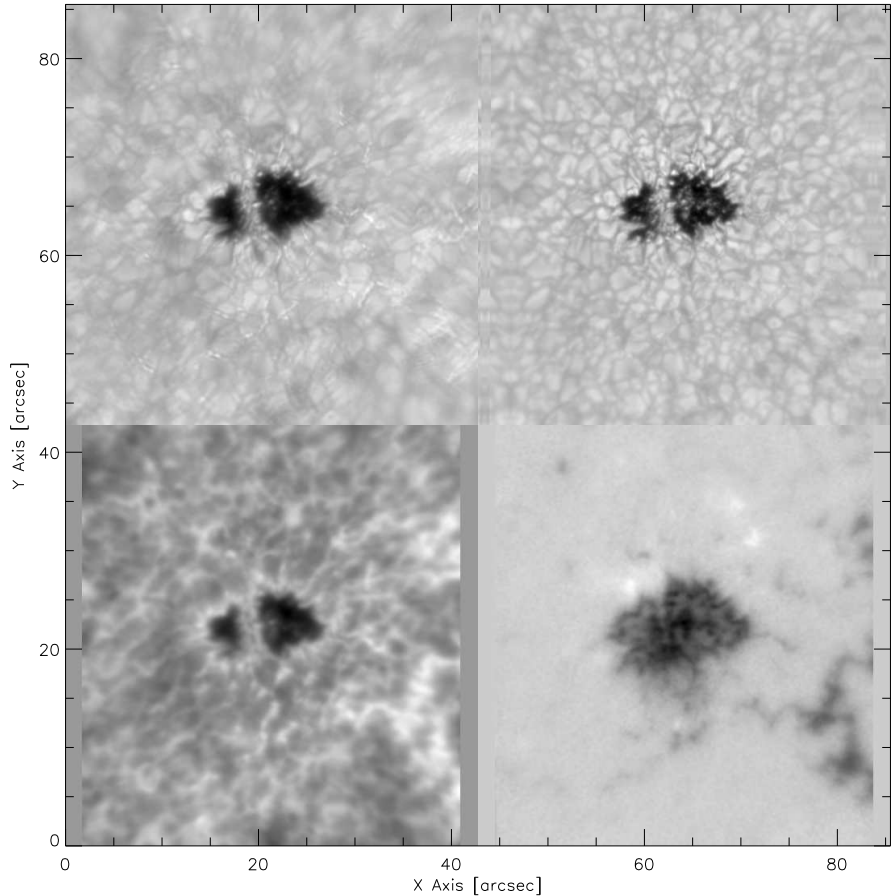}
   \end{tabular}
   \end{center}
   \caption 
   { \label{fig:2008} 
Some sample images from the 2008 data-set data-cubes. Upper-left: G-band image; Upper-right: Broad-band image; Lower-left: Stokes-I image near the core of the FeI 617.3 nm line; Lower-right: Stokes-V image near the wing of the FeI 617.3 nm line. For the sake of visualization, all images have been rescaled to the same pixel resolution (0.167" pixel$^{-1}$), their values have been separately linearly scaled to saturate the greyscale palette and we only show the central part of the Broad-band and spectropolarimetric images.}   
   \end{figure*} 
The second data-set used \cite{criscuoli12,sobotka12} in this work was acquired on $15$th October $2008$, at the DST.
The region imaged is AR $11005$ which manifested as a small pore with a light bridge in the northern hemisphere of the Sun at a high latitude [25.2$^{\circ}$ N, 10.0$^{\circ}$ W].\\
The spectropolarimetric part of the data-set consists of $80$ repetition of a $21$-point-scan of the FeI $617.3$ nm line by IBIS.
The difference in wavelength between the sampled points of the FeI line was $0.002$ nm and the exposure time for each image was set to $80$ ms. Each spectral scan took approximately $52$ s to complete.
The pixel scale of these $512 \times 512$ images is $0.167$ arcsec.
The six modulation states I $\pm$ S have been acquired with the following temporal scheme: S = [+Q , +V , -Q, -V, -U, +U].\\
For each spectropolarimetric image, we also acquired a broad-band white light ($621.3 \pm 5$ nm) counterpart, imaging the same field-of-view (a sample is shown in the upper-right panel of Fig. \ref{fig:2008}).
The pixel scale of the $1024 \times 1024$ broad-band image was set at $0.083"$ and the exposure time was $80$ ms (shared shutter with IBIS spectropolarimetric images). 
The broad-band images have been restored with MFBD \cite{MFBD}, returning a single wide-band frame for comparison to each scan.\\
As in the 2006 data-set case, the spectropolarimetric images have been registered and de-stretched to minimize the seeing effects uncorrected by the adaptive optics system and to achieve the highest spatial resolution.
After this process, the original IBIS spectropolarimetric $I \pm S$ images were successively combined and corrected for instrumental polarization to retrieve the Stokes vectors (samples of Stokes I and Stokes V images are shown in the lower-left and lower-right panels of Fig. \ref{fig:2008}, respectively).
The estimated average spatial resolution of the spectropolarimetric scans used in this work is $0.36"$.\\
Also in this data-set, G-band filtergrams ($430.5\pm0.5$~nm) are present; the original pixel scale of such $1024\times1024$ pixel images was $0.037''$, while the exposure time was $10$~ms.
Again, a post processing MFBD procedure was applied to those images, which were afterward re-binned at $4\times$ the spectropolarimetric spatial scale (a sample is shown in the upper-left panel of Fig. \ref{fig:2008}).\\
For the compression analysis, the spectropolarimetric and broad-band image were cropped to their $256 \times 512$ central parts, while in this case the G-band filtergrams already had a $1024 \times 1024$ size (as shown on Fig. \ref{fig:2008} and resumed in Tab. \ref{tab:2008}).
All the data-cubes were saved as 32bit greyscale.\\
\begin{table*}[h]
\begin{small}
\caption{2008 data-set description}
\label{tab:2008}       
\begin{tabular}{lccccc}
\hline\noalign{\smallskip}
Type & Image Size & N. of frames & Repetitions & Spatial Resolution & Pixel scale \\
\noalign{\smallskip}\hline\noalign{\smallskip}
G-band & 1024$\times$1024 & 3  & 80 & $0.10"$ & $0.04"$\\ 
\hline 
Broad-band & 256$\times$256 & 21 &  80 & $0.33"$ & $0.167"$ \\ 
\hline 
Stokes I & 256$\times$256 & 21 & 80 & $0.36"$ & $0.167"$ \\ 
\hline 
Stokes V & 256$\times$256 & 21 & 80 & $0.36"$ & $0.167"$\\ 
\noalign{\smallskip}\hline
\end{tabular}
\end{small}
\end{table*}


\section{Results and Discussion}
Here we present the results of the compression and decompression tests, which have been performed on an AMD Athlon 64 X2 Dual Core Processor 4400+ with 8GB RAM, 750GB Hitachi Deskstar 7K1000 hard drive and 64bit Linux Fedora 23 Operating System.\\
All the test have been performed on 32bit greyscale data-cubes, unless where explicitly written otherwise.\\
The compression tests were performed using the default values for the JP3D encoder.
In brief, the resolution levels computed per volume dimension through the DWT are 3 for the spatial dimensions and 1 for the wavelength or time dimension. The code block size is $64\times64\times64$,  the size of the precincts at each resolution are $2^{15} \times 2^{15} \times 2^{15}$ and no sub-sampling has been applied. We refer to Part 10 of the JPEG2000 standard \cite{JPEG2000,github_Readme} for an in-deep explanation of the parameters of the JP3D compression algorithm.\\
The results are resumed in tables and ordered by the type of data-cube used, comparing the two different data-set and different approaches in the presentation of the data-cube to the encoder.
For each data-type, the data-cube size before compression, the average, minimum and maximum CR (Compression Ratio) obtained, the average BPV (Bit per Voxel) needed to compress the data and the average times to compress and to de-compress the data-cube are presented.
For the sake of visualization, these information have been presented in two separated tables.\\

\subsection{Broad-band data-cubes}
\begin{table*}[h]
\begin{small}
\caption{Broad-band data-cubes compression ratios}
\label{tab:BB1}       
\begin{tabular}{lcccc}
\hline\noalign{\smallskip}
Description & Size [MB] & CR AVE& CR min& CR MAX\\
\noalign{\smallskip}\hline\noalign{\smallskip}
\textbf{2006 Data}& & & & \\ 
Broad-band x,y,t [256,256,270] &  68.0  & 5.7 & 5.2 & 9.9\\
{16bit} Broad-band x,y,t [256,256,270] &  34.0 & 2.7 & 2.6 & 3.0\\
\textbf{2008 Data}& & & & \\ 
Broad-band x,y,t [256,512,21] &  10.5 & 8.5 & 6.9 & 10.6\\
{16bit} Broad-band x,y,t [256,512,21] & 5.2 & 4.6 & 3.9 & 5.3\\
\noalign{\smallskip}\hline
\end{tabular}
\end{small}
\end{table*}

\begin{table*}[h]
\begin{small}
\caption{Broad-band data-cubes BPV and Times}
\label{tab:BB2}       
\begin{tabular}{lcccc}
\hline\noalign{\smallskip}
Description & Size [MB] & BPV AVE& t Comp [s]& t Decomp [s]\\
\noalign{\smallskip}\hline\noalign{\smallskip}
\textbf{2006 Data}& & & & \\ 
Broad-band x,y,t [256,256,270] &  68.0 & 5.7 & 25.0 & 32.0\\
{16bit} Broad-band x,y,t [256,256,270] & 34.0 & 6.0 & 26.4 & 33.4\\
\textbf{2008 Data}& & & & \\ 
Broad-band x,y,t [256,512,21] & 10.5 & 3.8 & 3.1 & 4.4\\
{16bit} Broad-band x,y,t [256,512,21] & 5.2 & 3.5 & 3.1 & 4.4\\
\noalign{\smallskip}\hline
\end{tabular}
\end{small}
\end{table*}
\textbf{Difference between the Broad-band data-cubes:\\}
2006: Quiet Sun granulation in the FOV.
We have 50 data-cubes, 68MB each, with dimension $256\times256\times270$ voxels.
Dimension x and y are spatial, z is time.
Spatial sampling is $\approx0.18"$, temporal sampling is $\approx0.3$ s.
Seeing extremely good; homogeneous image quality in the whole FOV.\\
2008: a pore in the FOV.
We have 80 data-cubes, 10.5MB each, with dimension $256\times256\times21$ voxels.
Dimension x and y are spatial, z is time.
Spatial sampling is $\approx0.17"$, temporal sampling is $\approx2.5$ s.
Seeing very good at image centre (AO lock point), degrading rapidly towards the periphery of the FOV.

\textbf{Discussion:\\}
All the raw images which were used to compose the data-cubes were acquired as 16bit images. After the calibration and restoring pipelines, they were all saved as 32bit data-cubes (as stated in Par. \ref{sect:data}) to be compliant with floating point representation standards \cite{IEEE754}.
As first test, we compressed the same broad-band data-cubes saved as 16bit and 32bit, to verify if the type of format of the data could affect the performances of the JP3D encoder or in the BPV needed to describe the data.\\
%
As reported in Tables \ref{tab:BB1} and \ref{tab:BB2}, the time needed to compress or decompress the 32bit or the 16bit version of the same data-cube is the same with very good approximation.
Also the average BPV needed to encode the data-cubes is essentially the same.\\
We interpret this last result as an evidence of the correct performance of the JP3D encoding.\\

As reported in Tables \ref{tab:BB1} and \ref{tab:BB2}, the 32bit version of the 2006 data-set can be compressed on average by a factor 5.7, the 32bit version of the 2008 data-set by a factor 8.5.
There are some cubes in both cases that reach a CR as high as 10, but are relative to the moments of very bad seeing which are present at the end of both data-sets.
In those moments, the contrast of the images drops and the data-cubes information content is poor.
The times to compress and decompress are not strictly proportional to the data-cube size, probably there is some correction due to the higher CR of the 2008 data.\\
It is worth to remark that the CR of the 2008 data-set is significantly higher.
This is probably due to the higher information content in the 2006 images: those images have extremely high contrast in the whole FOV, while the 2008 data-set contrast is very high only the centre of the FOV and decreases rapidly as we get closer and closer to the periphery.
Apparently, there is no gain in sampling at a higher frequency the evolution of the solar structure: sampling at 2.5 s or 0.3 s with a pixel scale of $\simeq0.17"$ is in any case more than enough to sample the evolution of the solar structure (at $\simeq0.17"$ we can sample features moving at sound speed with $\simeq10$ s cadence).\\
%

To resume: the 2006 Broad-band data-cubes needed $\approx$6 BPV on average, while the 2008 Broad-band data-cubes needed $\approx$4 BPV on average.
These results have to be compared with the 8 BPV needed when compressing single Broad-band images with JPEG2000 \cite{delmoro11}.
Also, the difference in encoding images saved in 16bit or 32bit is negligible.
\subsection{G-band data-cubes}
\begin{table*}[h]
\begin{small}
\caption{G-band data-cubes compression ratios}
\label{tab:GB1}       
\begin{tabular}{lcccc}
\hline\noalign{\smallskip}
Description & Size [MB] & CR AVE& CR min& CR MAX\\
\noalign{\smallskip}\hline\noalign{\smallskip}
\textbf{2006 Data}& & & & \\ 
G-band x,y,t [1024,1024,18] &  72.0  & 10.3 & 9.1 & 14.9\\
G-band x,y,t {BIG} [1024,1024,288] &  1200.0 & 9.5 & * & *\\
G-band x,y,t {RE-BIN} [512,512,18] &  18.0 & 6.7 & 6.5 & 7.5\\
\textbf{2008 Data}& & & & \\ 
G-band x,y,t [1024,1024,3] &  12.0 & 12.0 & 9.7 & 14.7\\
G-band x,y,t {BIG} [1024,1024,240] & 960.0 & 10.3 & * & *\\
\noalign{\smallskip}\hline
\end{tabular}
\vspace{3 mm}
\\
* Since in this case we have only one large data-cube, its CR value is reported under CR AVE and CR min and CR MAX are left empty.
\\
\end{small}
\end{table*}

\begin{table*}[h]
\begin{small}
\caption{G-band data-cubes BPV and Times}
\label{tab:GB2}       
\begin{tabular}{lcccc}
\hline\noalign{\smallskip}
Description & Size [MB] & BPV AVE& t Comp [s]& t Decomp [s]\\
\noalign{\smallskip}\hline\noalign{\smallskip}
\textbf{2006 Data}& & & & \\ 
G-band x,y,t [1024,1024,18] &  72.0  & 3.2 & 23.0 & 35.0\\
G-band x,y,t {BIG} [1024,1024,288] &  1200.0 & 3.3 & 370.0 & 568.0\\
G-band x,y,t {RE-BIN} [512,512,18] &  18.0 & 4.8 & 7.3 & 10.6\\
\textbf{2008 Data}& & & & \\ 
G-band x,y,t [1024,1024,3] &  12.0 & 2.7 & 2.7 & 3.9\\
G-band x,y,t {BIG} [1024,1024,240] & 960.0 & 3.1 & 268.0 & 376.0\\
\noalign{\smallskip}\hline
\end{tabular}
\end{small}
\end{table*}
\textbf{Difference between the G-band data-cubes:\\}
2006: Quiet Sun granulation in the FOV.
We have 50 data-cubes, 72MB each, with dimension $1024\times1024\times18$ voxels.
Dimension x and y are spatial, z is time.
Spatial sampling is $\approx0.05"$, temporal sampling is $\approx5$ s.
Seeing extremely good; homogeneous image quality in the whole FOV.\\
2008: a pore in the FOV.
We have  80 data-cubes, 12MB each, with dimension $1024\times1024\times3$ voxels.
Dimension x and y are spatial, z is time.
Spatial sampling is $\approx0.04"$, temporal sampling is $\approx17.3$ s.
Seeing very good at image centre (AO lock point), degrading rapidly towards the periphery of the FOV.\\

\textbf{Discussion:\\}
The G-band data-cubes have been used to explore the results of different strategies in ordering and compressing the data-cubes.\\
First, we compare the performances of JP3D when compressing the 2006 G-band images as 50 $1024\times1024\times18$ voxel (72MB) data-cubes versus a single large $1024\times1024\times288$ voxel (1.2GB) data-cube and the 2008 G-band images as 80 $1024\times1024\times3$ voxel (12MB) data-cubes versus a single large $1024\times1024\times240$ voxel (~1GB) data-cube.
Results resumed in Tab. \ref{tab:GB1} and \ref{tab:GB2}.
In the case where small data-cubes are compressed, the CR varies between 9 and 15, the highest compressibility usually associated to moments of bad seeing and the lowest to moments of good seeing.
The average CRs are 10 and 12 for the 2006 data-set and the 2008 data-set, respectively, again the lowest value is obtained for the data-set which had the best seeing during the observations.
These values indicate that the G-band voxels of these data-sets need ~3 bits to be represented without information loss.
The time needed to compress and to decompress scales with the dimension of the data-cube, with very little discrepancies.
Therefore, the algorithm seems to be very efficient in managing large data-cubes, up to the file sizes used in this experiment.\\
Then, we focus on the differences when compressing a Nyquist under-sampled or over-sampled data-set \cite{Kotelnikov33}.
We stress that the pixel size of the 2006 G-band images is $0.05"$, therefore it is sampling appropriately the image according to the Nyquist definition \cite{Grenander59}, since the spatial frequency cut-off set by the finite aperture of the 76 cm DST telescope at 430.5 nm (the central wavelength of the G-band) is $\sim 0.12"$.
If we re-bin the $1024\times1024$ images to $512\times512$ images, we reduce the image size by a $4\times$ factor, while loosing part of the information allowed by the Nyquist sampling.
Nevertheless, if the structures present in the data-cube have angular dimensions such that are already correctly sampled at this lower resolution, the compressibility of the original and re-bin data-cubes should be the same. 
Instead, the comparison of the values for the re-bin and original data-cubes reported in Tab. \ref{tab:GB1} and \ref{tab:GB2} show that, on average, we obtain a lower compression ratio for the re-binned images (CR $\sim7$ instead of the $\sim10$ we obtain for the original images).
This fact hints that, if we under-sample the image, the nearby pixels are less correlated and more bits are needed to describe the average pixel.
Of course, this spatial correlation between the original pixels is probably due to the finite extent of the telescope Point Spread Function, which is properly sampled by the 0.05 arcsec scale and not so by the re-binned $0.1"$ scale.
Last, we notice that the different temporal sampling of the 2006 and 2008 G-band images seem to have little or no part in CR results, probably masked by the other differences (the type of the structure imaged and/or the seeing).\\

To resume: the 2006 and 2008 G-band data-cubes needed $\approx$3 BPV on average.
This result has to be compared with the 6 BPV or more needed when compressing single G-band images with JPEG2000 \cite{delmoro11}.
Also, there is limited gain in compressing a single large data-cube with respect to several smaller data-cubes, containing the same data volume.
Last, sampling the images above or below the Nyquist spatial frequency has an effect on the CR much smaller than the increase of the data-cube size due to the re-sampling.

\subsection{Spectropolarimetric data-cubes} 
\begin{table*}[h]
\begin{small}
\caption{Spectropolarimetric data-cubes compression ratios}
\label{tab:SP1}       
\begin{tabular}{lcccc}
\hline\noalign{\smallskip}
Description & Size [MB] & CR AVE& CR min& CR MAX\\
\noalign{\smallskip}\hline\noalign{\smallskip}
\textbf{2006 Data}& & & & \\ 
Stokes I x,y,$\lambda$ [256,256,45] &  11.2  & 4.4 & 4.1 & 6.0\\
Stokes I x,y,t [256,256,50] &  12.5 & 4.3 & 4.1 & 5.1\\
Stokes V x,y,$\lambda$ [256,256,45] &  11.2  & 5.8 & 5.5 & 6.1\\
Stokes V x,y,t [256,256,50] &  12.5 & 5.8 & 5.5  & 6.5\\
Stokes I$\pm$V x,y,$\lambda$ [512,256,45] &  23.0  & 4.2 & 4.1 & 4.3\\
\textbf{2008 Data}& & & & \\ 
Stokes I x,y,$\lambda$ [256,512,21] &  10.5  & 7.1 & 6.4 & 8.2\\
Stokes I x,y,t [256,512,80] &  40.0 & 7.1 & 6.5 & 7.6\\
Stokes V x,y,$\lambda$ [256,512,21] &  10.5  & 7.8 & 7.2 & 8.7\\
Stokes V x,y,t [256,512,80] & 40.0 & 7.8 & 6.9 & 8.4\\
Stokes I$\pm$V x,y,$\lambda$ [512,512,21] &  21.0  & 6.6 & 6.2 & 7.0\\
\noalign{\smallskip}\hline
\end{tabular}
\end{small}
\end{table*}

\begin{table*}[h]
\begin{small}
\caption{Spectropolarimetric data-cubes BPV and Times}
\label{tab:SP2}       
\begin{tabular}{lcccc}
\hline\noalign{\smallskip}
Description & Size [MB] & BPV AVE& t Comp [s]& t Decomp [s]\\
\noalign{\smallskip}\hline\noalign{\smallskip}
\textbf{2006 Data}& & & & \\ 
Stokes I x,y,$\lambda$ [256,256,45] &  11.2  & 7.3 & 5.7 & 7.2\\
Stokes I x,y,t [256,256,50] &  12.5 & 7.3 & 6.3 & 8.2\\
Stokes V x,y,$\lambda$ [256,256,45] &  11.2  & 5.6 & 4.1 & 5.6\\
Stokes V x,y,t [256,256,50] &  12.5 & 5.5 & 5.0 & 6.9\\
Stokes I$\pm$V x,y,$\lambda$ [512,256,45] &  23.0  & 7.7 & 11.2 & 14.0\\
\textbf{2008 Data}& & & & \\ 
Stokes I x,y,$\lambda$ [256,512,21] &  10.5  & 4.5 & 4.2 & 6.5\\
Stokes I x,y,t [256,512,80] &  40.0 & 4.5 & 16.0 & 25.0\\
Stokes V x,y,$\lambda$ [256,512,21] &  10.5  & 4.1 & 3.9 & 5.8\\
Stokes V x,y,t [256,512,80] & 40.0 & 4.1 & 13.9 & 21.8\\
Stokes I$\pm$V x,y,$\lambda$ [512,512,21] &  21.0  & 4.9 & 8.7 & 13.0\\
\noalign{\smallskip}\hline
\end{tabular}
\end{small}
\end{table*}
\textbf{Difference between the spectropolarimetric data-cubes:\\}
2006: Quiet Sun granulation in the FOV.
We have 50 data-cubes, 11.2MB each, with dimension $256\times256\times45$ voxels both for Stokes I and Stokes V.
Dimension x and y are spatial, z is wavelength.
Spatial sampling is $\approx0.18"$, temporal sampling is $\approx89$ s.
Seeing extremely good; homogeneous image quality in the whole FOV.\\
2008: a pore in the FOV.
We have 80 data-cubes, 10.5MB each, with dimension $256\times512\times21$ voxels both for Stokes I and Stokes V.
Dimension x and y are spatial, z is wavelength.
Spatial sampling is $\approx0.17"$, temporal sampling is $\approx52$ s.
Seeing very good at image centre (AO lock point), degrading rapidly towards the periphery of the FOV.\\

\textbf{Discussion:\\}
The spectropolarimetric data allowed us to compare the outcomes of JP3D compression when ordering the data-sets in the standard [x,y,t] order, to create $\lambda$-homogeneous data-cubes against the [x,y,$\lambda$] order, to create data-cubes made of images which are as close as possible to the same time-stamp (i.e.: t-homogeneous within the time cadence).
Tab. \ref{tab:SP1} and \ref{tab:SP2} report the results of these compression runs.
We see that for both the 2006 and 2008 data-sets there is no substantial difference in the CRs when using  $\lambda$-homogeneous data-cubes or  t-homogeneous data-cubes: Stokes I always needs $\sim 7$ or $\sim 4$ BPV for the 2006 and 2008 data-sets, respectively;  Stokes V always needs $\sim 5$ or $\sim 4$ BPV for the 2006 and 2008 data-sets, respectively.
This results suggests that, given the sampling strategies of these two data-sets, the temporal evolution and the change in the object due to changes in the observation wavelength are of the same order and there is very little difference in describing the first or the second with a wavelet-like approach.
Also the times needed to compress and de-compress the data-cube scale according to the file-size only.
With the same data-sets, we were also able to compare the compressibility of the I and V Stokes images when compressed in their standard visualization after the data calibration (i.e.: as I and V Stokes separated images) against their I$\pm$V realization (closer to the way they are actually acquired).  
The results reported in Tab. \ref{tab:SP1} and \ref{tab:SP2} show that the BPV needed to represent the average pixel of the I$\pm$V data-cubes are slightly greater than the Stokes I BPV for both the 2006 and 2008 data-sets.\\ 

To resume: there is apparently no gain in ordering the data in $\lambda$-homogeneous data-cubes or $t$-homogeneous data-cubes.
Stokes I data-cubes needed $\approx$7 BPV or $\approx$4 BPV on average, respectively for the 2006 and 2008 data-sets. Stokes V data-cubes needed $\approx$5 BPV or $\approx$4 BPV on average, respectively for the 2006 and 2008 data-sets.
These results have to be compared with the 10 BPV or more needed when compressing single spectropolarimetric images with JPEG2000 \cite{delmoro11}.

\begin{landscape}
\subsection{Summary Table} 
\begin{table}[h]
\caption{Resume table}
\label{tab:resume}       
\begin{tabular}{lcccc>{\bfseries}ccc}
\hline\noalign{\smallskip}
Description & Size [MB] & CR AVE& CR min& CR MAX & BPV AVE& t Comp [s]& t Decomp [s]\\
\noalign{\smallskip}\hline\noalign{\smallskip}
\textbf{2006 Data}& & & &  & & & \\ 
Broad-band x,y, t [256,256,270] &  68.0  & 5.7 & 5.2 & 9.9 & 5.7 & 25.0 & 32.0\\
\textbf{16bit} Broad-band x,y, t [256,256,270] &  34.0 & 2.7 & 2.6 & 3.0 & 6.0 & 26.4 & 33.4\\
G-band x,y,t [1024,1024,18] &  72.0  & 10.3 & 9.1 & 14.9 & 3.2 & 23.0 & 35.0\\
G-band x,y,t {BIG} [1024,1024,288] &  1200.0 & 9.5 & * & * & 3.3 & 370.0 & 568.0\\
G-band x,y,t {RE-BIN} [512,512,18] &  18.0 & 6.7 & 6.5 & 7.5 & 4.8 & 7.3 & 10.6\\
Stokes I x,y,$\lambda$ [256,256,45] &  11.2  & 4.4 & 4.1 & 6.0 & 7.3 & 5.7 & 7.2\\
Stokes I x,y,t [256,256,50] &  12.5 & 4.3 & 4.1 & 5.1 & 7.3 & 6.3 & 8.2\\
Stokes V x,y,$\lambda$ [256,256,45] &  11.2  & 5.8 & 5.5 & 6.1 & 5.6 & 4.1 & 5.6\\
Stokes V x,y,t [256,256,50] &  12.5 & 5.8 & 5.5  & 6.5 & 5.5 & 5.0 & 6.9\\
Stokes I$\pm$V x,y,$\lambda$ [512,256,45] &  23.0  & 4.2 & 4.1 & 4.3& 7.7 & 11.2 & 14.0\\
\textbf{2008 Data}& & & &  & & &  \\
Broad-band x,y,t [256,512,21] &  10.5 & 8.5 & 6.9 & 10.6 & 3.8 & 3.1 & 4.4\\
\textbf{16bit} Broad-band x,y,t [256,512,21] & 5.2 & 4.6 & 3.9 & 5.3& 3.5 & 3.1 & 4.4\\
G-band x,y,t [1024,1024,3] &  12.0 & 12.0 & 9.7 & 14.7 & 2.7 & 2.7 & 3.9\\
G-band x,y,t {BIG} [1024,1024,240] & 960.0 & 10.3 & * & * & 3.1 & 268.0 & 376.0\\
Stokes I x,y,$\lambda$ [256,512,21] &  10.5  & 7.1 & 6.4 & 8.2 & 4.5 & 4.2 & 6.5\\
Stokes I x,y,t [256,512,80] &  40.0 & 7.1 & 6.5 & 7.6 & 4.5 & 16.0 & 25.0\\
Stokes V x,y,$\lambda$ [256,512,21] &  10.5  & 7.8 & 7.2 & 8.7 & 4.1 & 3.9 & 5.8\\
Stokes V x,y,t [256,512,80] & 40.0 & 7.8 & 6.9 & 8.4 & 4.1 & 13.9 & 21.8\\
Stokes I$\pm$V x,y,$\lambda$ [512,512,21] &  21.0  & 6.6 & 6.2 & 7.0 & 4.9 & 8.7 & 13.0\\
\noalign{\smallskip}\hline
\end{tabular}
\vspace{3 mm}
\\
* Since in this case we have only one large data-cube, its CR value is reported under CR AVE and CR min and CR MAX are left empty.
\\
\end{table}
\end{landscape}

\section{Conclusions}
\label{sect:conclusion}
The present generation of solar spectral imagers generates large data volumes, which are difficult to transfer and needs large storage capacities.
The upcoming 4m class of solar observatories are foreseen to acquire and store data at a rate larger of one or two order of magnitudes than present time.
The problem of storing and transmitting such large volumes can be mitigated with suitable algorithms.\\
In this paper, we studied the performances of JP3D for lossless compression on to calibrated data-cubes of different types.
All the data-set were acquired at NSO-Sac-Peak and then reduced and calibrated (using IBIS pipeline, and/or MFMBD and/or de-stretching).
These data-sets are a partial description of the present variety of high resolution photospheric solar data, which will probably be maintained in the next years.\\
To summarize the results of our analysis: a) The performance of the 3D compression varies with the data-type: G-band are the most compressible (~3 bits per voxel), while Stokes I are the less compressible (~7 bits per voxel);
b) The CR is about twice as large as in the case of the 2D JPEG2000 compression;
c) The gain in ordering data in [x,y,$\lambda$,t] or [x,y,t,$\lambda$] is apparently negligible;
d) The algorithm seems to be efficient in handling the larger files, with little differences in compressing a single large data-cube or several smaller data-cubes; 
e) The spatial correlation present in data which are super-sampled with respect to the telescope cut-off frequency leads to a +33$\%$ in compression rate.
Obviously, that super-sampling required 4 times the number of voxel, leading to 4$\times$ the initial data volume.\\
We also note that, even considering an enhancement of the data processing capabilities in the foreseeable future, the compression time will be a crucial factor.
In fact, the acquisition rates are likely to increase up to $\sim 100 \; 1024\times1024$ pixel images per second and real time compression of the data-sets may not be feasible.
This has to be taken in consideration in the observatory daily schedule.\\
To conclude, the excellent performances in lossless compression of the JP3D algorithm suggest to explore also its performances in lossy compression, in those cases where the preservation of all the information is not critical.\\

\begin{acknowledgements}
This study has been partially supported by the SOLARNET project (www.solarnet-east.eu), funded by the European Commission's FP7 Capacities Programme under the Grant Agreement 312495.
\end{acknowledgements}

\bibliographystyle{spmpsci}
\bibliography{report}

\end{document}